\documentclass[aps,amsmath,amssymb,prb,twocolumn]{revtex4}

\usepackage{graphicx}
\usepackage{times}
\usepackage{amsmath}
\usepackage{amsfonts}
\usepackage{amssymb}
\usepackage{units}
\DeclareGraphicsExtensions{.pdf,.png,.eps,.jpg}

\begin{document}
\title{A Guide to the Design of Electronic Properties of Graphene Nanoribbons}

\author{Oleg V. Yazyev }
\affiliation{Institute of Theoretical Physics, Ecole Polytechnique F\'ed\'erale de Lausanne (EPFL), CH-1015 Lausanne,
Switzerland}

\date{\today}

\begin{abstract}

{\it Conspectus: }
Graphene nanoribbons (GNRs) are one-dimensional nanostructures predicted to display a rich variety of electronic behaviors.  Depending on their structure, GNRs realize metallic and semiconducting electronic structures with band gaps that can be tuned across broad ranges. Certain GNRs also exhibit a peculiar gapped magnetic phase for which the half-metallic state can be induced as well as the topologically non-trivial quantum spin Hall electronic phase. Because their electronic properties are highly tunable, GNRs have quickly become a popular subject of research toward the design of graphene-based nanostructures for technological applications. This Account presents a pedagogical overview of the various degrees of freedom in the atomic structure and interactions that researchers can use to tailor the electronic structure of these materials. The Account provides a broad picture of relevant physical concepts that would facilitate the rational design of GNRs with desired electronic properties through synthetic techniques.

We start by discussing a generic model of zigzag GNR within the tight-binding model framework.  We then explain how different modifications and extensions of the basic model affect the electronic band structures of GNRs.  We classify the modifications based on the following categories:  (1) electron-electron and spin-orbit interactions, (2) GNR configuration, which includes width and the crystallographic orientation of the nanoribbon (chirality), and (3) the local structure of the edge. We  subdivide this last category into two groups: the effects of the termination of the π-electron system and the variations of electrostatic potential at the edge.  This overview of the structure-property relationships provides a view of the many different electronic properties that GNRs can realize.

The second part of this Account reviews three recent experimental methods for the synthesis of structurally well-defined GNRs.  We describe a family of techniques that use patterning and etching of graphene and graphite to produce GNRs.  Chemical unzipping of carbon nanotubes also provides a route towards producing chiral GNRs with atomically smooth edges.  Scanning tunneling microscopy/spectroscopy investigations of these unzipped GNRs have revealed edge states and strongly suggest that these GNRs are magnetic.  The third approach exploits the surface-assisted self-assembly of GNRs from molecular precursors.  This powerful method can provide full control over the atomic structure of narrow nanoribbons and could eventually produce more complex graphene nanostructures.

\end{abstract}

\maketitle

\section{Introduction}

The discovery of graphene, a two-dimensional allotrope of carbon, has opened new horizons in both basic science and technology.\cite{1,2,3,4,5,6,7} The ever-growing interest in this nanomaterial originates from several sources. To start with, graphene is the first truly two-dimensional material, which makes it an ideal platform for exploring two-dimensional physics. Graphene is a semi-metal with low-energy quasiparticles being described by a Dirac-like equation. The electronic structure of graphene gives rise to a number of extraordinary physical properties, such as the exceptionally high charge-carrier mobilities, making this material very interesting due to possible technological applications in the domain of electronics. Finally, novel electronic properties of graphene can be tailored in a number of conceptually new ways.

One approach for controlling electronic properties of graphene exploits the idea of producing one-dimensional nanometers-wide strips of graphene, commonly referred to as graphene nanoribbons (GNRs). Since the seminal theory work of Nakada {\it et al.},\cite{8} GNRs have been attracting considerable attention. Their striking feature is the diversity of electronic structure patterns that can be obtained by changing the structural parameters, like width and crystallographic orientation of the edges. For instance, whilst electronic band gaps of GNRs cut along the armchair lattice direction depend sensitively on the nanoribbon width, zigzag GNRs exhibit the presence of non-bonding edge states that were predicted to realize a peculiar type of magnetic ordering.\cite{8,9,10,11,12,13,14} The magnetically ordered state of zigzag GNRs can be controlled by means of an external electric field,\cite{15} which can be exploited in spintronics, the extension of traditional electronic utilizing electron spins for storing and manipulating information. The list of ``tuning knobs'' for controlling electronic properties of GNRs investigated theoretically is not limited to their width and edge orientation. Local atomic structure of the edges turns out to be equally important. However, progress on the experimental side of this direction of research has been hindered by the lack of synthetic routes for producing structurally well-defined GNRs. This situation only started to change very recently.

\renewcommand{\figurename}{Scheme}
\begin{figure*}
\includegraphics[width=160mm]{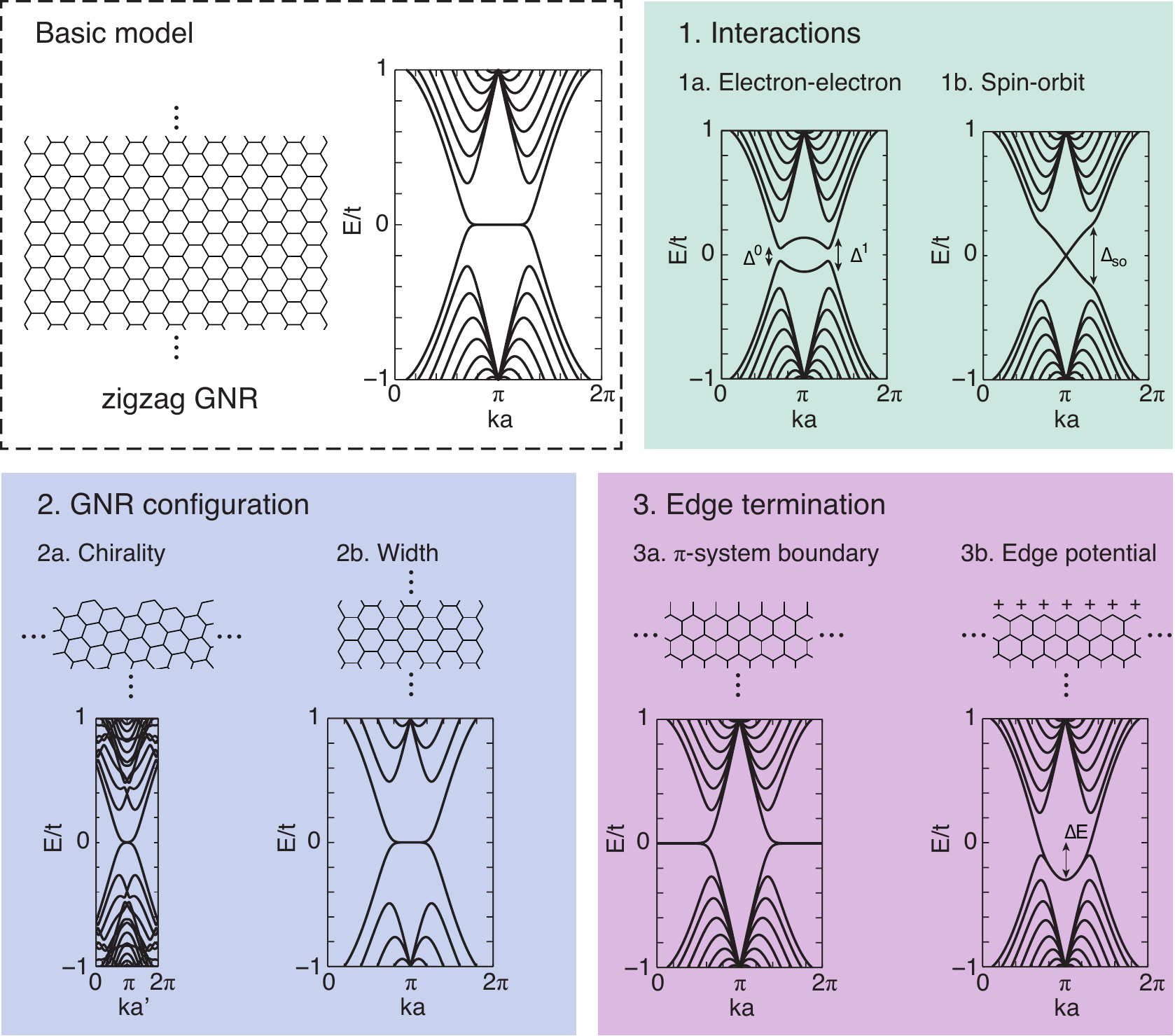}
\caption{
Classification of different extensions of the basic model – zigzag GNR described by the tight-binding model Hamiltonian. The scheme distinguishes interactions (electron-electron and spin-orbit), GNR configuration (chirality and width) and edge termination ($\pi$-electron system boundary and edge potential). Atomic structures and band structure plots are shown for each extension of the model.
}
\label{sch1}
\end{figure*}

This Account aims at presenting a unified overview of various degrees of freedom that can be used for tailoring the electronic structure of GNRs. The discussion is built upon a theoretical model employing the tight-binding Hamiltonian applied to the structure of zigzag GNR. We then consider different extensions of the basic theoretical model organized in the following manner: (1) interactions, (2) GNR configuration and (3) edge termination. The first category considers physical interactions not accounted for by the basic tight-binding model. This includes electron-electron interactions described by the on-site Coulomb repulsion term, a necessary ingredient for describing the magnetically ordered state of zigzag and chiral GNRs, as well as weak spin-orbit interactions that might be responsible for the emergence of topologically non-trivial electronic phases in graphene. Second category ``GNR configuration'' includes the structural parameters of GNR as a whole: the already mentioned width and edge orientation. ``Edge termination'' refers to the atomic-scale details of GNR edges. This category is further subdivided into the effects of $\pi$-electron system termination and the variations of local electrostatic potential at the edge.  This bird’s eye view of the structure-property relations serves to facilitate the rational design of GNRs with desired electronic properties by a broad community of experimental researchers. 

The second part of this Account reviews the most recent advances in producing well-defined GNRs. In particular, the following three approaches are considered. A family of techniques based on patterning and etching graphene and graphite for producing GNRs is described in the first part. Second, chemical unzipping of carbon nanotubes provides a route towards producing chiral GNRs with atomically smooth edges. The scanning tunneling microscopy/spectroscopy investigations of GNRs obtained using this method reveal the presence of edge states, as well as the strong indication of their magnetic nature. The third approach exploits the surface-assisted self-assembly of GNRs starting from molecular precursors. This powerful method is capable of providing full control over the atomic structure of narrow nanoribbons and can potentially be used for producing more complex graphene nanostructures.

\section{Theoretical model}

{\bf Tight-binding model.} The basic physical model underlying the present discussion of the electronic structure of graphene is the tight-binding model. The model considers only the $\pi$-symmetry electronic states formed by the unhybridized $p_z$ atomic orbital of $sp^2$ carbon atoms in graphene. The tight-binding model Hamiltonian can be written as
\begin{equation}
{\mathcal H}=-t\sum_{\left<i,j\right>,\sigma} \left[ c_{i\sigma}^{\dagger}c_{j\sigma} + c_{j\sigma}^{\dagger}c_{i\sigma} \right ]
\label{eq1}
\end{equation}
where the operators  $c_{i\sigma}$ ($c_{i\sigma}^{\dagger}$) annihilate (create) an electron with spin $\sigma$ at atom $i$. Here we restrict our consideration to nearest-neighbor pairs of atoms $\left<i,j\right>$. The only empirical parameter in this model is the hopping integral $t \approx 2.7$~eV, which is assumed to be independent of external conditions. The described tight-binding model written in the physicist’s language is fully equivalent to the H\"uckel method more familiar to chemists. Surprisingly, this extremely simple model is able to describe correctly, and often accurately, the electronic structure of graphene systems in most situations. 

The starting point for discussing the structure-property relations in GNRs is a generic structural model of zigzag GNR, shown in the head panel of Scheme~\ref{sch1}. The dots indicate the periodicity direction of the nanoribbon. Only the $sp^2$ carbon atoms belonging to the $\pi$-electron system are shown for clarity. The dangling bonds at the edge are generally assumed to be passivated by some functional groups that do not couple to the $\pi$-electron system and have electronegativity similar to that of carbon atoms, e.g. hydrogen atoms. 

\renewcommand{\figurename}{Figure}
\begin{figure}
\includegraphics[width=82.5mm]{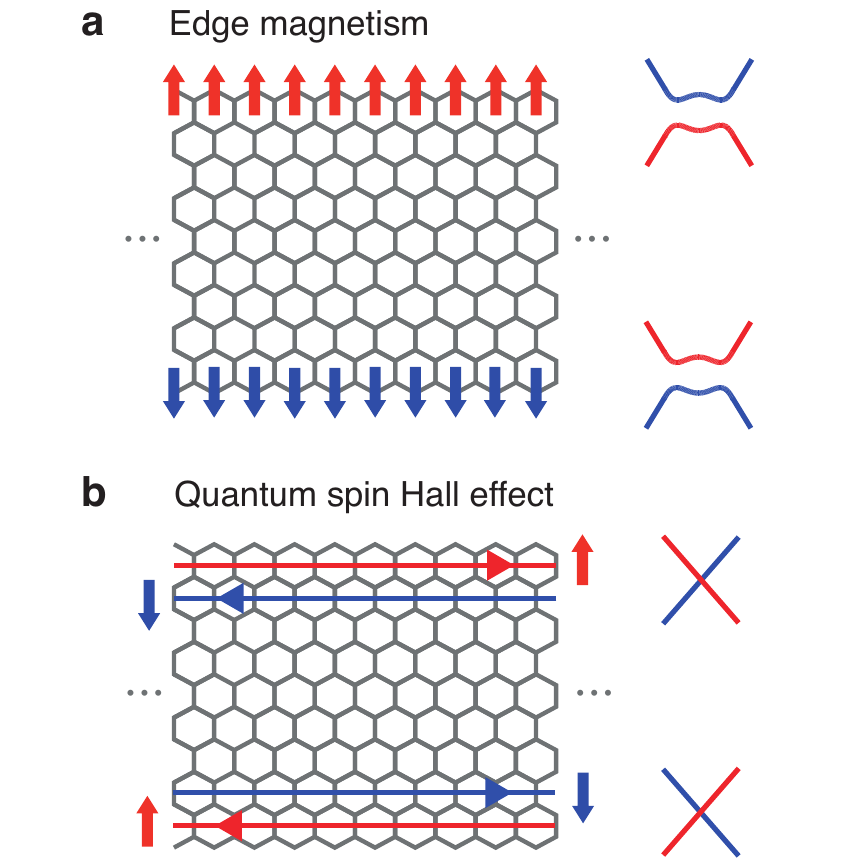}
\caption{
Schematic drawing of (a) the edge magnetism and (b) the quantum spin Hall effect in zigzag GNRs. The schematic edge-resolved band structures (right) are colored according to the spin direction of the electronic states involved.
}
\label{fig1}
\end{figure}

The electronic band structure calculated using the tight-binding method is shown next to the structural model of zigzag GNR in Scheme~\ref{sch1}. The band structure plot spans the complete one-dimensional Brillouin zone of the nanoribbon. By convention, the range of momenta $0 < k < 2\pi/a$ is shown ($a = 0.246$~nm is the lattice constant of zigzag GNR). Without loss of generality, the energy is shown in units of hopping integral $t$. In this band structure one can recognize a series of sub-bands derived from the two Dirac cones of the two-dimensional band structure of ideal graphene, located at $k = 2\pi/3a$ and $k = 4\pi/3a$. However, a new feature is the flat band at $E = 0$ and $2\pi/3a < k < 4\pi/3a$. This flat band corresponds to non-bonding electronic states localized at the edges, a common property of zigzag GNRs of all widths, which is closely related to the bipartite symmetry of the crystalline lattice of graphene. 

{\bf Electron-electron interactions.} The presence of the flat band feature gives rise to the high density of electronic states at the Fermi level. The associated electronic instability can be relieved by the onset of magnetic ordering driven by the electron-electron interactions. In order to describe these effects one would need electron-electron interactions to be accounted for explicitly in the Hamiltonian. This can be achieved by including the Hubbard term 
\begin{equation}
{\mathcal H_{\rm e-e}}=U\sum_i n_{i\uparrow} n_{i\downarrow} 
\label{eq2}
\end{equation}
where $n_{i\sigma} = c_{i\sigma}^{\dagger} c_{i\sigma}$ is the spin-resolved electron density at atom $i$. The empirical parameter in the Hubbard term is the strength of on-site Coulomb repulsion $U$, which was estimated to be of the same order of magnitude as $t$.\cite{16} This justifies employing the mean-field approximation to the Hubbard term\cite{17,18}
\begin{equation}
{\mathcal H_{\rm e-e}^{\rm mf}}= U\sum_i \left ( n_{i\uparrow} \left < n_{i\downarrow} \right > + \left < n_{i\uparrow} \right > n_{i\downarrow} - \left < n_{i\uparrow} \right > \left < n_{i\downarrow} \right > \right )
\label{eq3}
\end{equation}
in which $\left < n_{i\sigma} \right >$ are the mean spin-resolved electron populations. The mean-field approximation greatly simplifies the calculation, essentially making this model Hamiltonian approach equivalent to the lattice version of the unrestricted Hartree-Fock method. 

The band structure of the basic zigzag GNR model calculated using the mean-field Hubbard model at $U = t$ is shown in Scheme~\ref{sch1} (see panel ``1a. Electron-electron interactions''). The introduction of electron-electron interactions lifts the degeneracy of edge states, opening a band gap. There are two characteristic splittings in the band structure: the band gap $\Delta^0$ and the zone-boundary splitting $\Delta^1$ at $k = \pi/a$. The electron-electron interactions give rise to a peculiar type of magnetic ordering illustrated in Figure~\ref{fig1}a. In the ground state magnetic moments are localized at the edges: the correlations are ferromagnetic along the edge and antiferromagnetic across the nanoribbon. Son, Cohen and Louie suggested that the gap in this ordered system can be closed selectively in either spin channel by applying a transverse electric field, thus making the system half-metallic.\cite{15} If realized in practice this effect would allow for the electrical control of spin transport with potential applications in spintronics. However, one has to appreciate the specifics of carbon-based magnetism compared to the traditional magnetic materials based on $d$-elements. Firstly, due to the weak spin-orbit interactions in graphene discussed below, the magnetic anisotropy of carbon-based magnetic systems is very low. Secondly, the magnetic ordering at the edges is one-dimensional. Strictly speaking, this implies that there is no long-range magnetic ordering at finite temperatures, but the system is characterized by a certain finite spin correlation length.\cite{19} The estimated spin correlation length at the zigzag edges of graphene is only of the order of one nanometer at room temperature, but can, in principle, achieve micrometer magnitudes at temperatures below 10~K.\cite{20}

\begin{figure}[b]
\includegraphics[width=82.5mm]{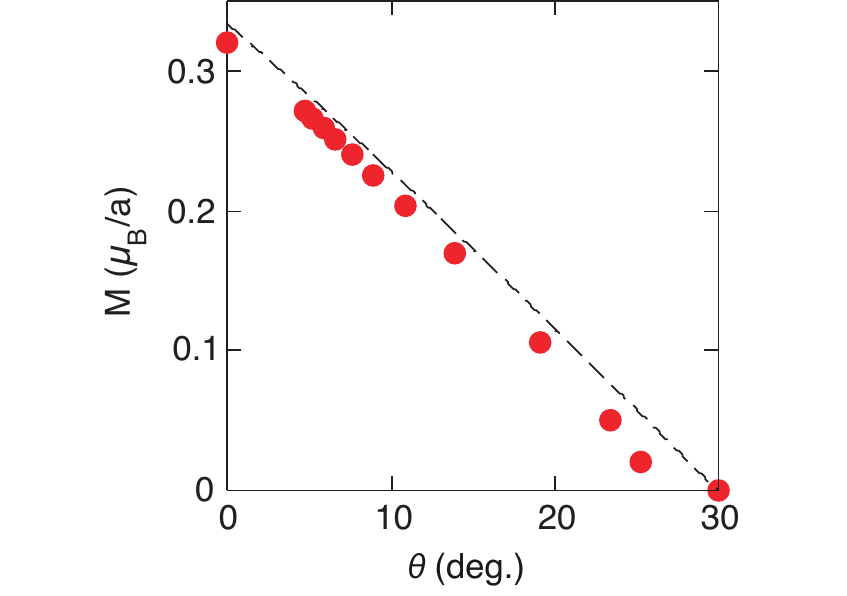}
\caption{
Magnetic moment per edge unit length $M$ as a function of chirality angle $\theta$ (circles) calculated for approximately 7-nm-wide GNRs using the mean-field Hubbard model at $U/t = 1$. The dashed line corresponds to the edge-state density $n(\theta)$ in the limit of infinite width.
}
\label{fig2}
\end{figure}

{\bf Spin-orbit interactions.} The spin-orbit interaction is an effect of relativistic origin that couples electron spin and orbital momentum. In our model Hamiltonian description of graphene, spin-orbit effects can be introduced by means of a second-nearest-neighbor hopping term\cite{21} 
\begin{equation}
{\mathcal H_{\rm s-o}}= \frac{2i}{\sqrt{3}} t' \sum_{\left < \left < i,j \right > \right >} c_i^\dagger \sigma \cdot \left ( {\mathbf d}_{kj} \times {\mathbf d}_{ik} \right ) c_j 
\label{eq4}
\end{equation}
in which the corresponding creation and annihilation operators are written in the two-component form (e.i. $c_i^\dagger = (c_{i\uparrow}^\dagger,c_{i\downarrow}^\dagger)$). In this expression, $\left < \left < i,j \right > \right >$ refers to the pairs of second-nearest-neighbor atoms $i$ and $j$ connected via atom $k$, $\sigma$ are the Pauli matrices, vector ${\mathbf d}_{ik}$ points from $k$ to $i$. The strength of spin-orbit interaction is given by empirical parameter $t'$. Spin-orbit interactions open a band gap $\Delta_{\rm so}$ at the Dirac points and lift the degeneracy of edge states in graphene (see Scheme~\ref{sch1}, $t' = 0.05t$ in the band structure plot). In the presence of spin-orbit interactions the edge states in graphene exhibit the properties of the quantum spin Hall effect boundary states. That is, spin-up quasiparticles propagate in one direction along the edge, while spin-down quasiparticles propagate in the opposite direction (Fig.~\ref{fig1}b). The consequence of such spin-momentum coupling is the absence of backscattering of conduction edge channels, as well as the intrinsic relation between charge current and spin density. The prediction of the quantum spin Hall effect in graphene by Kane and Mele\cite{21} was one of the milestone works in the development of the emerging field of topological insulators,\cite{22,23,24} which are typically heavy-element materials. In graphene, however, the involved energy scales are too small because of the very weak relativistic spin-orbit coupling. First-principles calculations predict that the magnitude of the spin-orbit gap in graphene is of the order of 10$^{-5}$~eV.\cite{25} No experimental confirmation of the quantum spin Hall effect in graphene has been reported so far. However, recipes for enhancing spin-orbit interaction on graphene by means of the heavy-element modification have been suggested.\cite{26}

{\bf Nanoribbon chirality.} Chirality refers to the orientation of the edges of GNRs with respect to the crystalline lattice of graphene. Chirality is commonly characterized by the chirality angle $\theta$. Zigzag and armchair high-symmetry directions correspond to $\theta = 0^\circ$ and $\theta = 30^\circ$, respectively, while chiral directions correspond to all intermediate values of $\theta$. The tight-binding band structures of GNRs show systematic dependence on the chirality angle. As $\theta$ increases the two Dirac cones of graphene band structure projected onto the one-dimensional Brillouin zone of GNR approach each other. The length in momentum space of the flat edge-state band connecting the two Dirac points decreases, and, therefore, the density of edge states per edge unit length also diminishes.\cite{27} This is illustrated in Scheme~\ref{sch1} for the specific case of $\theta=19.1^\circ$ chiral GNR. Note that the scale of this band-structure plot accounts for different one-dimensional Brillouin-zone dimensions as a consequence of larger periodicity length  ($a' = 6.51$~\AA) compared to the case of zigzag GNR ($a = 2.46$~\AA). When electron-electron interactions are included the edge states undergo complete spin-polarization. The density of edge states translates directly into the magnetic moment per edge unit length (Fig.~\ref{fig2}). In this plot the circles correspond to magnetic moments per edge unit length $M$, calculated for approximately 7-nm-wide GNRs using the mean-field Hubbard model ($U/t = 1$). The line corresponds to the density of edge states per unit length
\begin{equation}
n(\theta) = \frac{2}{3a} \cos \left(\theta + \frac{\pi}{3} \right)
\label{eq5}
\end{equation}
in the limit of infinite GNR width.\cite{28}

The case of armchair GNRs ($\theta = 30^\circ$,) is somewhat special, since this is the only edge orientation which preserves the equivalence of the two sublattices of the bipartite lattice of graphene. The edge states vanish in this case, as follows from Expr.~(\ref{eq5}). All armchair GNRs are either semiconductors or metals with bulk-like states at the charge-neutrality point. 

\begin{figure}
\includegraphics[width=82.5mm]{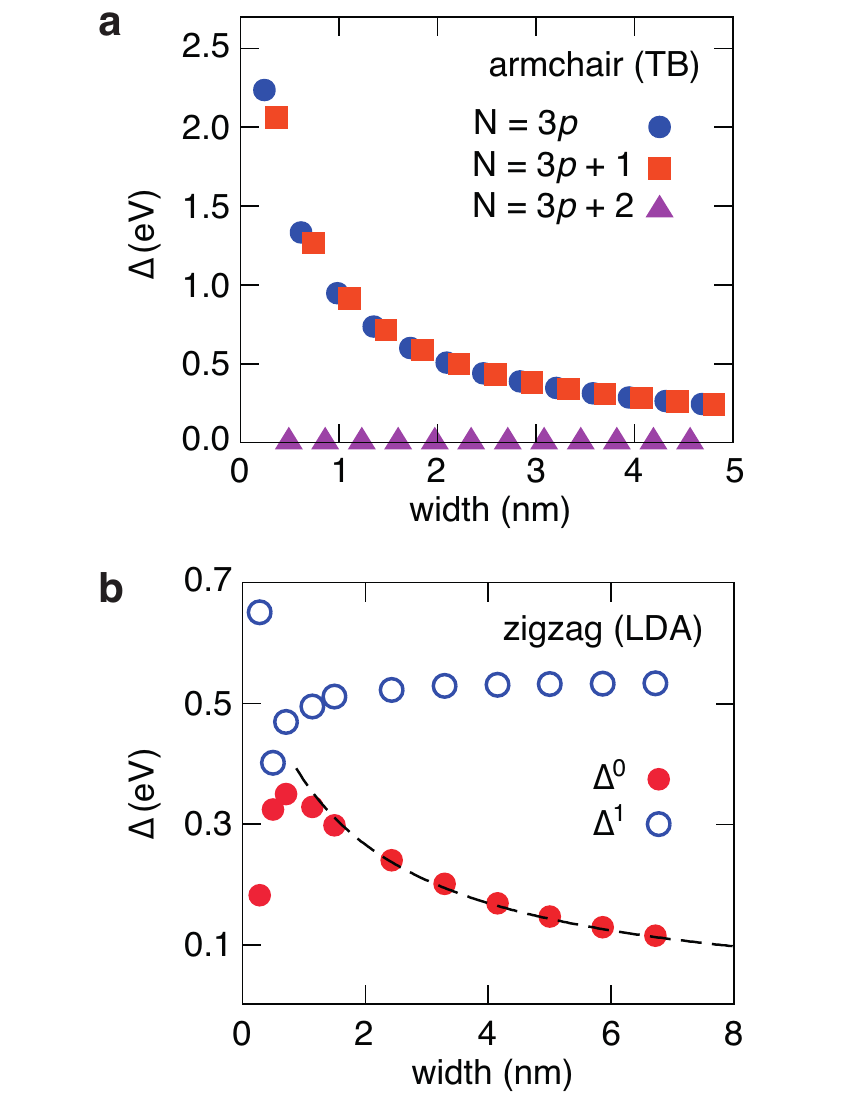}
\caption{
(a) Band gaps of armchair GNRs as a function width calculated using the tight-binding (TB) approach. Three families of armchair GNRs are distinguished. (b) Band gaps $\Delta^0$ and zone-boundary splittings $\Delta^1$ of the spin-polarized grounds states of zigzag GNRs as a function of their width calculated using local density approximation (LDA). Adapted with permission from Ref.~\onlinecite{12}. Copyright 2006 American Physical Society. 
}
\label{fig3}
\end{figure}

{\bf Nanoribbon width.} The width of GNRs is another important descriptor of the overall configuration of a GNR. Scheme~\ref{sch1} shows the tight-binding band structure of a zigzag GNR when the width is a factor of two smaller compared to the initial structure. The edge-state band does not change, but the reduction of width leads to a larger spacing between the bulk sub-bands as a result of quantum confinement. The band gaps of armchair GNRs, however, display a very systematic dependence on width (Fig.~\ref{fig3}a). Within the tight-binding picture the armchair GNRs are divided into three families according to their electronic structure: semiconducting (families $N = 3p$ and $N = 3p + 1$; $N$ refers to the number of pairs of atoms per GNR unit cell), and metallic ($N = 3p + 2$). First-principle calculations predict that $N = 3p + 2$ armchair GNRs are semiconductors as well. The band gaps $\Delta$ of semiconducting armchair GNRs are predicted to be inversely proportional to the width within one family. The overall relation is an oscillating dependence dumped by 1/width decay. A similar inversely proportional dependence show the band gaps ($\Delta^0$) of the spin-polarized gapped phases of zigzag and chiral GNRs in the presence of the Hubbard term in the model Hamiltonian (Fig.\ref{fig3}b). Thus, the 1/width behavior of band gaps of GNRs is a common feature irrespective of their origin (either quantum confinements as in armchair GNRs or spin-polarization in the case of zigzag and chiral GNRs).

Besides chirality and width, the local structural details of the edge play an equally important role in defining the electronic properties of GNRs. Here it is convenient to distinguish between two contributions to the local edge structure: the topology of the $\pi$-electron system boundary and the local electrostatic environment at the edge. 

\renewcommand{\figurename}{Scheme}
\begin{figure}[b]
\includegraphics[width=82.5mm]{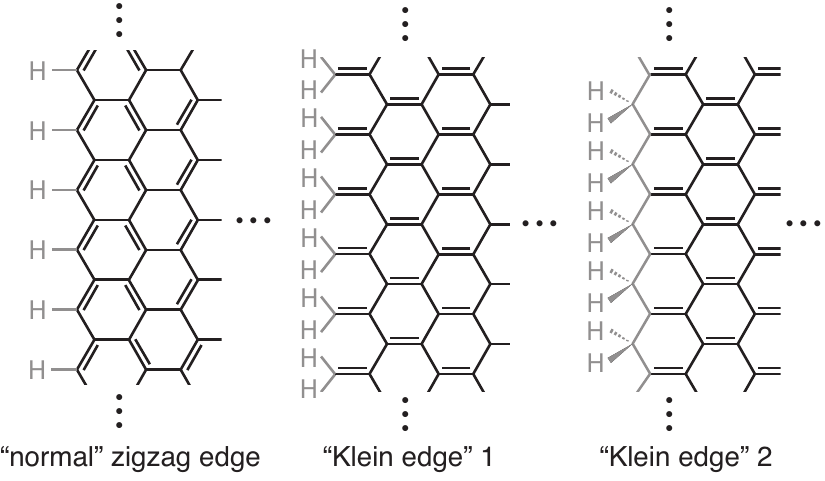}
\caption{
Three different possible hydrogen terminations of graphene edge oriented along the zigzag direction leading to the ``normal'' and the ``Klein edge'' $\pi$-electron system boundaries. Covalent bonds belonging to the $\pi$-electron system are highlighted with darker lines.
}
\label{sch2}
\end{figure}

{\bf $\pi$-System termination.} Until now we were assuming the most straightforward edge termination, in which the outmost $sp^2$ carbon atoms have two nearest neighbors in the π-electron system of graphene. However, other possibilities do exist, e.g. the so-called ``Klein edge'' zigzag termination, in which the outmost $sp^2$ carbon atoms are bonded to only one carbon atom in the $\pi$-electron system.\cite{29} Starting from the initially considered ``normal'' zigzag termination, the ``Klein edge'' termination can be realized in two different ways: by extending the edge with methylidene groups or by hydrogenating the edge carbon atoms, which effectively results in their removal from the $\pi$-electron system due to the rehybidization into the $sp^3$ state (Scheme~\ref{sch2}).  The tight-binding band structure of the ``Klein edge'' terminated zigzag GNR shows the presence of an edge-state flat band (Scheme~\ref{sch1}). However, the band now connects the two Dirac points by passing through the Brillouin zone center ($k = 0$). This effectively doubles the length flat-band segment in momentum space, and, therefore, doubles the density of edge states. More complex edge terminations may result from self-passivation relieving the structural instability of bare zigzag edges that were predicted theoretically\cite{30} and later observed in transmission electron microscopy experiments.\cite{31,32} Importantly, certain $\pi$-electron system termination topologies may result in dramatic changes of the electronic structure, e.g. leading to the complete quenching of edge states at the zigzag edges.\cite{33}

{\bf Edge potential.} The effects of the local electrostatic environment are probably the easiest to understand. Potential changes at the edge can originate from the difference in electronegativities between carbon atoms and the functional groups terminating the edge, as well as from other factors, such as adsorbates or interactions with the substrate on which the GNR is deposited. These effects can be modeled by adding the following term to the Hamiltonian
\begin{equation}
{\mathcal H_{\rm pot}}= \sum{i,\sigma} \epsilon_i c_{i\sigma}^\dagger c_{i\sigma}
\label{eq6}
\end{equation}
in which on-site potentials $\epsilon_i$ are imposed at the edge. In general, the edge potential results in either upward or downward bending of the flat band. In illustration shown in Scheme~\ref{sch1}, the potential is set to $−0.03t$ for the outmost edge atoms. This modification resulted in an edge-state band bending by $\Delta E =  −0.03t$. 

\section{Experimental progress}

Whilst theoretical understanding of the electronic properties of graphene nanostructures has advanced considerably in the past few years, the production of well-defined GNRs has been demonstrated only recently. Below we review three main approaches capable of producing GNRs with atomically precise edges. 

\renewcommand{\figurename}{Figure}
\begin{figure}[b]
\includegraphics[width=82.5mm]{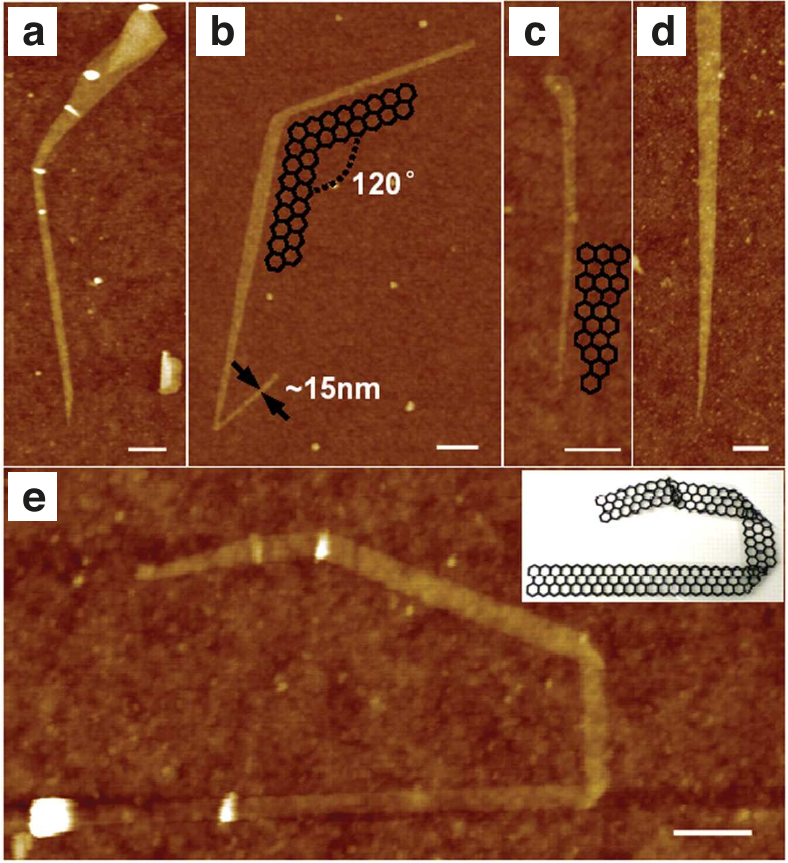}
\caption{
Atomic force microscopy images of chemically derived GNRs showing a variety of morphologies. All scale bars are 100 nm. Reproduced with permission from Ref.~\onlinecite{38}. Copyright 2008 American Association for the Advancement of Science.
}
\label{fig4}
\end{figure}

{\bf Etching graphene and graphite.} In their pioneering work Han {\it et al.} produced GNRs from mechanically exfoliated graphene by e-beam patterning followed by oxygen plasma etching.\cite{34} This top-down approach allowed for controlled productions of GNRs with sub-50-nm width while lacking the possibility of controlling edge orientation and structure. The patterned GNRs, however, displayed a clear semiconducting behavior and rather efficient field-effect transistor devices were demonstrated. A subsequent modification of the original process employing nanowires as an etch mask allowed to reduce the GNRs width down to 6~nm.\cite{35} In both cases, the band gaps of GNRs showed an inverse proportional dependence on width as expected from theoretical models discussed in the previous sections. A somewhat different technique exploits the catalytic etching of graphene by thermally activated Ni nanoparticles.\cite{36,37} Such catalytic etching is highly anisotropic and allows to ``cut'' graphene along high-symmetry crystallographic orientations. Moreover, certain control over the direction (armchair or zigzag) was achieved via the size of catalytic nanoparticles. Sub-10-nm width GNRs as well as more complicated graphene nanostructures have been obtained using this catalytic etching technique. 

The group of Hongjie Dai introduced a solution-based technique allowing to produce GNRs starting from bulk graphite.\cite{38,39} The first stage of the process is the intercalation of graphite with sulfuric and nitric acids followed by rapid heating to 1000$^\circ$C. Expanded graphite is then sonicated in organic solved with a suitable surfactant to yield a stable suspension of GNRs with widths down to 10~nm (Fig.~\ref{fig4}). Electrical transport measurements revealed that all GNRs produced using this procedure are semiconducting with band gaps inversely proportional to the width and achieving the values as high as 0.4~eV at 10~nm. The investigated GNRs displayed hole mobilities of 100-−200~cm$^2$~V$^{−-1}$~s$^{−-1}$ and transistor behavior with on/off current ratios approaching 10$^7$ at room temperature.

\begin{figure}
\includegraphics[width=85mm]{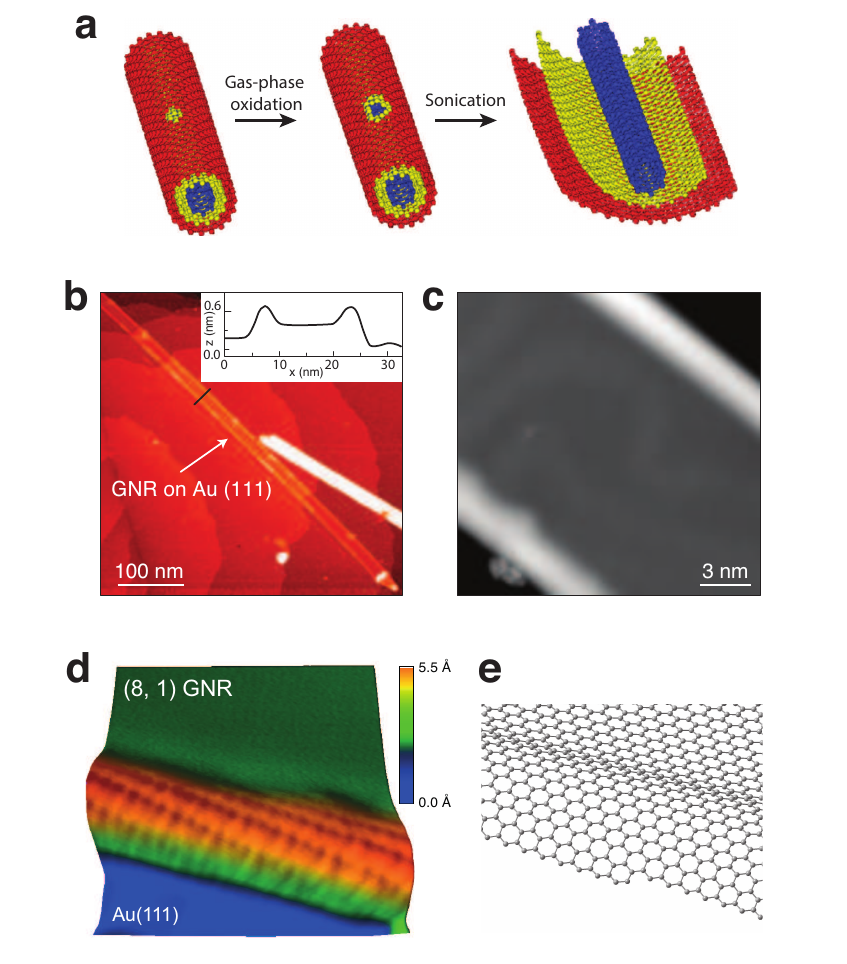}
\caption{
(a) Schematic illustration of the unzipping of multi-walled carbon nanotubes via the two-stage process involving gas-phase oxidation damage and solution-phase sonication. Reproduced with permission from Ref.~\onlinecite{42}. Copyright 2010 Macmillan Publishers Limited. (b) STM image of monolayer GNR on Au(111) surface. The inset shows the line profile indicated by a black line. (c) Higher resolution STM image of a GNR at $T = 7$~K. (d) Atomically resolved STM image of the edge of an (8,1) GNR and (e) the corresponding structural model. Reproduced with permission from Ref.~\onlinecite{43}. Copyright 2011 Macmillan Publishers Limited. 
}
\label{fig5}
\end{figure}

{\bf Unzipping carbon nanotubes.} High-quality GNRs can be produced by several techniques that involve the chemical unzipping of carbon nanotubes.\cite{40,41,42} The basic methodology for one of these unzipping techniques\cite{42} is illustrated in Fig.~\ref{fig5}a. First, the precursor multiwalled carbon nanotubes are locally damaged by oxidation in air at 500$^\circ$C. Then, mechanical sonication in an organic solvent ``unzips'' the nanotube along its axis, resulting in straight GNRs of uniform width. The GNRs obtained in this way inherit the width and chirality of the precursor nanotubes. Due to the broad distribution of both the diameter and chirality of carbon nanotubes, the resulting GNRs are generally chiral with practically random edge orientations, and are characterized by a relatively broad distribution of width (10--20~nm). 

\begin{figure}
\includegraphics[width=85mm]{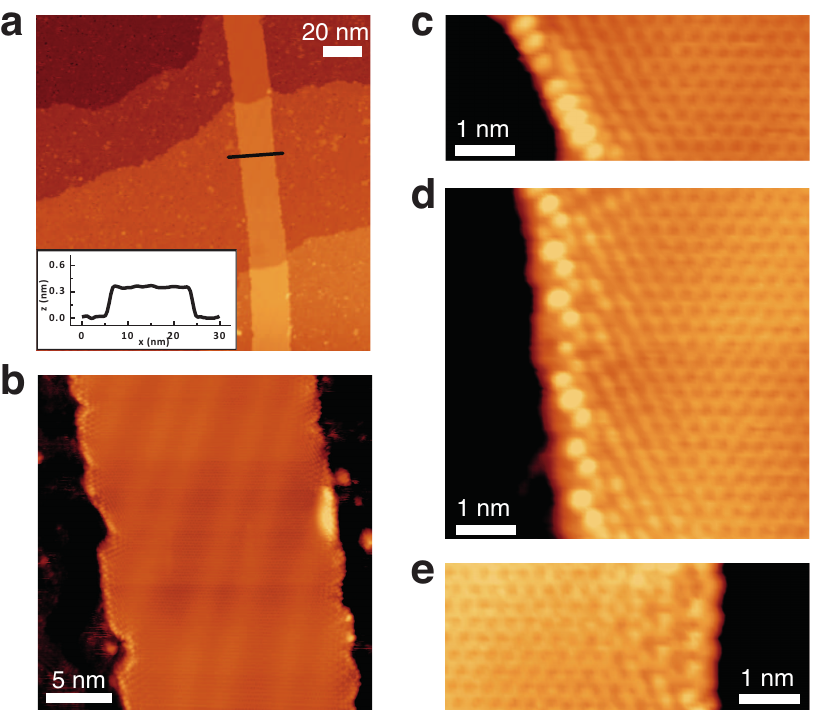}
\caption{
(a) Room-temperature STM image of a GNR on Au(111) surface after hydrogen plasma treatment. The inset shows the line profile along the black line. (b) Magnified STM image of a hydrogen plasma treated GNR reveals the rough morphology of the edges, as well as the presence of edge states. (c-−e) Atomically resolved STM images of zigzag, chiral ($\theta = 19^\circ$) and armchair straight segments of a graphene edge, respectively. Reproduced with permission from Ref.~\onlinecite{45}. Copyright 2012 American Chemical Society.
}
\label{fig6}
\end{figure}

\begin{figure*}
\includegraphics[width=165mm]{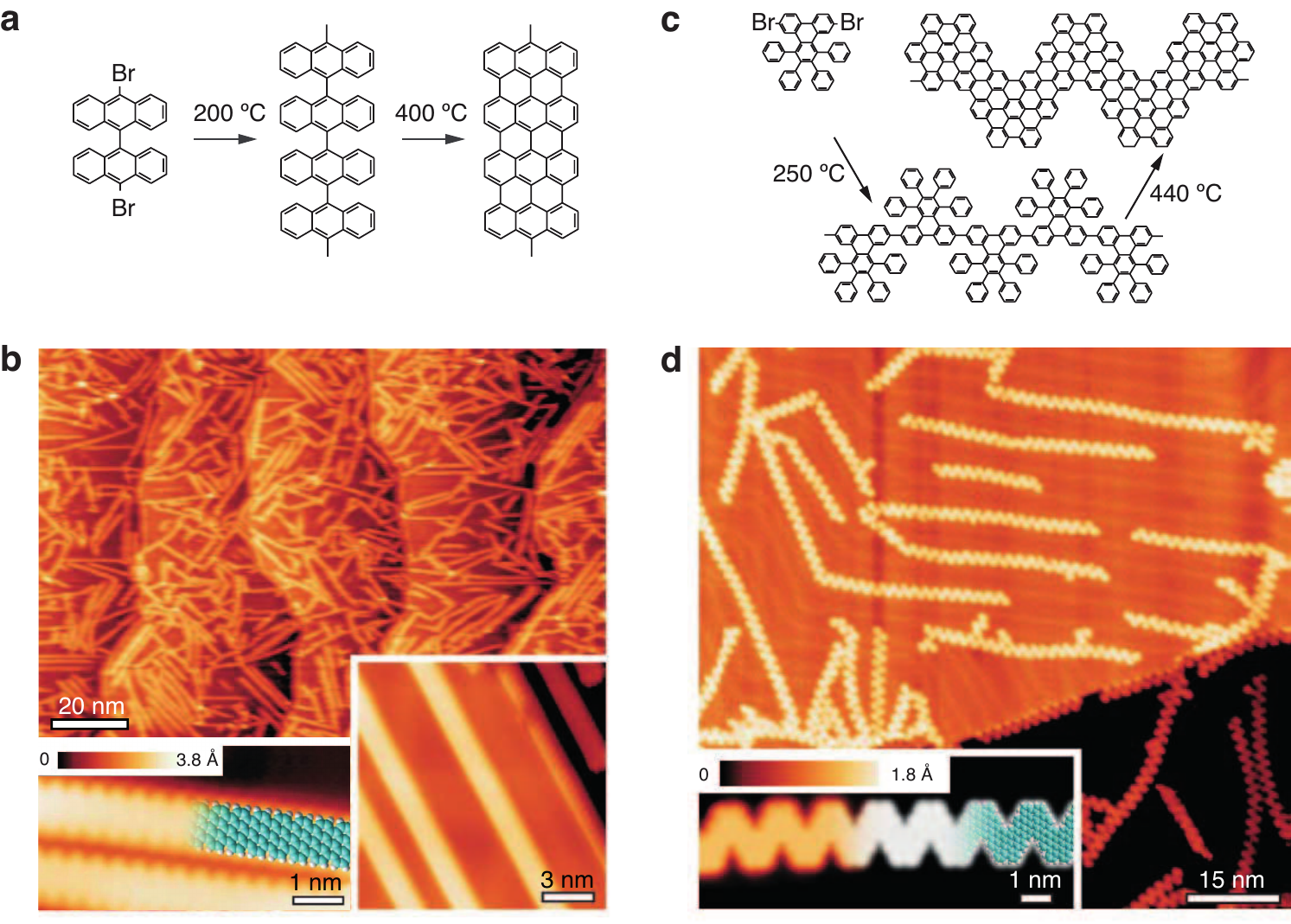}
\caption{
(a) Reaction scheme of the formation of narrow armchair GNRs starting from molecular precursor 10,10’-dibromo-9,9’-bianthryl ({\bf 1}). (b) STM images of the synthesized armchair GNRs on Au(111) surface recorded at different resolutions. (c) Reaction scheme of the synthesis of chevron-type GNRs starting from 6,11-dibromo-1,2,3,4-tetraphenyltriphenylene ({\bf 2}). (d) STM images of the chevron-type GNRs on Au(111) surface. Reproduced with permission from Ref.~\onlinecite{46}. Copyright 2010 Macmillan Publishers Limited.
}
\label{fig7}
\end{figure*}

The electronic structure of GNRs produced via the nanotube unzipping route was experimentally investigated by Tao {\it et al.} using scanning tunneling microscopy (STM) and spectroscopy (STS) techniques.\cite{43} High-resolution STM images revealed the atomically smooth edges of these GNRs and allowed determination of their chirality (Fig.~\ref{fig5}b-e). The STS measurements show the presence of electronic states localized at the edges in the vicinity of the Fermi level for all chiral GNRs. Importantly, the edge states appear to be split into a pair of peaks separated by 20−-50 meV. This splitting was interpreted as a strong indication of the spin-polarization of the edge states.  This interpretation is corroborated by the model Hamiltonian calculations,\cite{27} which show the same dependence of the STS peak splittings on the chirality and width of nanoribbons. Similar STS signatures have recently been observed in GNRs synthesized using the chemical vapor deposition technique.\cite{44}

Even though atomic resolution was achieved in the experiments described above, STM is unable to reveal the details of edge termination. This is due to the non-zero contact angle between the edge of the GNR and the Au substrate used in the experiments. The strong GNR-substrate interactions possibly originate from the presence of oxygen-containing functional groups at the edge. In more recent work Zhang {\it et al.} have shown that hydrogen plasma etching of the GNRs flushes the undesired functional groups, leaving hydrogen-terminated edges that lie flat on the substrate.\cite{45} The edges of etched GNRs are no longer straight, but this allows the observation of few-nanometers-long edge segments with different chiralities (zigzag, armchair and chiral) within the same GNR (Fig.~\ref{fig6}). The study confirms the presence of edge state at both chiral and zigzag edges. The combination of STM observations and first-principles calculations shows that the edges of hydrogen-plasma-etched GNRs are free of structural reconstructions (i.e. show no non-six-membered rings) and are terminated by hydrogen atoms with no rehybridization of the outermost edge carbon atoms (i.e. one hydrogen atom saturates one edge carbon atom). 

{\bf Bottom-up approaches.} The bottom-up self-assembly route starting from molecular precursors is able to provide a high degree of control over the structure of narrow ($\sim$1~nm) GNRs. A two-step process introduced by Cai {\it et al.} is illustrated in Figure~\ref{fig7}a.\cite{46,47} The molecular precursors are 10,10’-dibromo-9,9’-bianthryl monomers ({\bf 1}) deposited on Au(111) surface, which form linear chains after dehalogenation at 200$^\circ$C. Further increase of temperature leads to cyclodehydrogenation resulting in armchair GNRs with well defined width and edge structure, as confirmed by the STM images shown in Figure~\ref{fig7}b. Recent STS and photoemission spectroscopy measurements performed on these armchair GNRs deposited on Au surface reported a band gap of 2.3$\pm$0.1~eV in good agreement with results of calculations employing many-body perturbation theory corrections.\cite{48} The power of this synthetic approach lies in the fact that it is possible to design molecular precursors leading to uniquely defined GNR structures. An example of a one-dimensional graphene nanostructure obtained starting from another ``bivalent'' precursor, 6,11-dibromo-1,2,3,4-tetraphenyltriphenylene ({\bf 2}), is shown in Figures~\ref{fig7}c,d. The resulting structure is a chevron-type GNR with locally armchair edges. Using molecular precursors capable of coupling to more than 2 units opens endless possibilities for designing multi-terminal graphene junctions, the necessary building blocks for designing complex nanometer-scale electronic circuits based on graphene. 

\begin{figure}
\includegraphics[width=82.5mm]{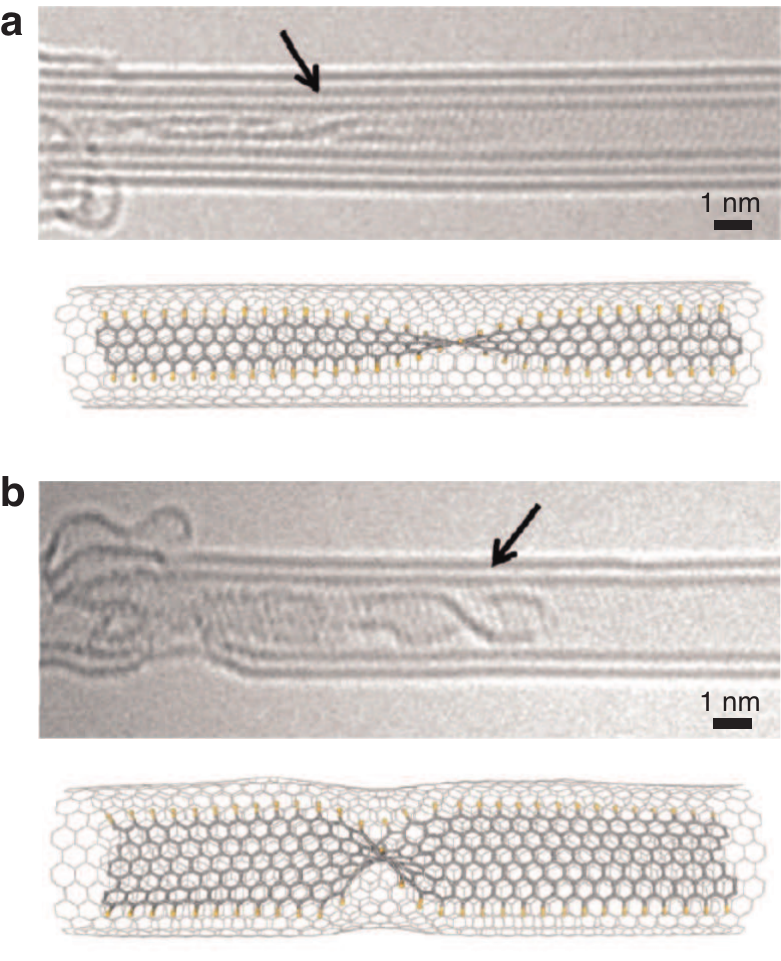}
\caption{
(a,b) High-resolution transmission electron microscopy images of sulphur-terminated zigzag GNRs of different widths confined inside multi-walled carbon nanotubes. Corresponding structural models are shown. Reproduced with permission from Ref.~\onlinecite{50}. Copyright 2012 American Chemical Society.
}
\label{fig8}
\end{figure}

So far the surface self-assembly method has been able to produce GNRs with locally armchair edges. A complementary approach exploits self-assembly inside carbon nanotube templates starting from sulfur-rich organic precursors such as tetrathiafulvalene (TTF) under heating ($>$800$^\circ$C) or subject to electron beam irradiation.{\cite{49,50} The width of sulfur-terminated zigzag GNRs obtained using this technique correlates with the diameter of the nanotube template. Two examples of transmission electron microscopy (TEM) images of sulfur-terminated GNRs inside carbon nanotubes are shown in Figure~\ref{fig8}.

\section{Summary and Outlook}

The Account explained in a pedagogical manner how different electronic structure patterns – metals and semiconductors with variable band gaps, magnetic and quantum spin Hall electronic phases – emerge in graphene nanoribbons. The presented model description provides a common view that intends to facilitate the rational design of graphene nanoribbons with desired electronic properties. Combined with the recent advances in controlled production of graphene nanoribbons reviewed in this Account – graphene etching techniques, carbon nanotube unzipping and the bottom-up chemical self-assembly – the described theoretical framework shall advance future developments aiming towards the technological application of graphene-based nanostructures.

\vskip 0.5cm

{\bf Acknowledgments.} This work was supported by the Swiss National Science Foundation (Grant No. PP00P2\_133552).

\noindent
{\bf Biographical information.} Oleg V. Yazyev received his M.Sc. degree in Chemistry from Moscow State University, Russia in 2003 and Ph.D. degree in Chemistry and Chemical Engineering from Ecole Polytechnique F\'ed\'erale de Lausanne (EPFL), Switzerland in 2007. Afterwards he spent several years as a postdoctoral fellow in the Department of Physics at the University of California, Berkeley and at the Lawrence Berkeley National Laboratory. In 2011 he started an independent research group at the Institute of Theoretical Physics, EPFL, supported by the Swiss National Science Foundation professorship grant and by an ERC Starting grant. His current research focuses on the theoretical and computational physics of graphene and topological insulators, with strong emphasis on their prospective technological applications.


\begin{thebibliography}{50}

\bibitem{1}
Novoselov, K. S.; Geim, A. K.; Morozov, S. V.; Jiang, D.; Zhang, Y.; Dubonos, S. V.; Grigorieva, I. V.; Firsov, A. A.: Electric Field Effect in Atomically Thin Carbon Films. {\it Science} {\bf 2004}, 306, 666−--669.

\bibitem{2} 
Berger, C.; Song, Z.; Li, T.; Li, X.; Ogbazghi, A. Y.; Feng, R.; Dai, Z.; Marchenkov, A. N.; Conrad, E. H.; First, P. N.; de Heer, W. A.: Ultrathin Epitaxial Graphite: 2D Electron Gas Properties and a Route toward Graphene-based Nanoelectronics. {\it J. Phys. Chem. B} {\bf 2004}, 108, 19912--−19916.

\bibitem{3} 
Geim, A. K.; Novoselov, K. S.: The rise of graphene. {\it Nature Mater.} {\bf 2007}, 6, 183−--191.

\bibitem{4} 
Katsnelson, M. I.: Graphene: carbon in two dimensions. {\it Materials Today} {\bf 2007}, 10, 20--−27.

\bibitem{5}
Castro Neto, A. H.; Guinea, F.; Peres, N. M. R.; Novoselov, K. S.; Geim, A. K.: The electronic properties of graphene. {\it Rev. Mod. Phys.} {\bf 2009}, 81, 109--−162.

\bibitem{6}
Novoselov, K. S.: Nobel Lecture: Graphene: Materials in the Flatland. {\it Rev. Mod. Phys.} {\bf 2011}, 83, 837--−849.

\bibitem{7}
Geim, A. K.: Nobel Lecture: Random walk to graphene. {\it Rev. Mod. Phys.} {\bf 2011}, 83, 851−--862.

\bibitem{8}
Nakada, K.; Fujita, M.; Dresselhaus, G.; Dresselhaus, M. S.: Edge state in graphene ribbons: Nanometer size effect and edge shape dependence. {\it Phys. Rev. B} {\bf 1996}, 54, 17954−--17961.

\bibitem{9}
Fujita, M.; Wakabayashi, K.; Nakada, K.; Kusakabe, K.: Peculiar Localized State at Zigzag Graphite Edge. {\it J. Phys. Soc. Jpn.} {\bf 1996}, 65, 1920−--1923.

\bibitem{10}
Barone, V.; Hod, O.; Scuseria, G. E.: Electronic Structure and Stability of Semiconducting Graphene Nanoribbons. {\it Nano Lett.} {\bf 2006}, 6, 2748--−2754.

\bibitem{11}
Ezawa, M.: Peculiar width dependence of the electronic properties of carbon nanoribbons. {\it Phys. Rev. B} {\bf 2006}, 73, 045432.

\bibitem{12}
Son, Y.-W.; Cohen, M. L.; Louie, S. G.: Energy Gaps in Graphene Nanoribbons. {\it Phys. Rev. Lett.} {\bf 2006}, 97, 216803.

\bibitem{13}
Yazyev, O. V.; Katsnelson, M. I.: Magnetic Correlations at Graphene Edges: Basis for Novel Spintronics Devices. {\it Phys. Rev. Lett.} {\bf 2008}, 100, 047209.

\bibitem{14}
Yazyev, O. V.: Emergence of magnetism in graphene materials and nanostructures. {\it Rep. Prog. Phys.} {\bf 2010}, 73, 056501.

\bibitem{15}
Son, Y.-W.; Cohen, M. L.; Louie, S. G.: Half-metallic graphene nanoribbons. {\it Nature} {\bf 2006}, 444, 347--−349.

\bibitem{16}
Yazyev, O. V.: Magnetism in Disordered Graphene and Irradiated Graphite. {\it Phys. Rev. Lett.} {\bf 2008}, 101, 037203.

\bibitem{17}
Feldner, H.; Meng, Z. Y.; Honecker, A.; Cabra, D.; Wessel, S.; Assaad, F. F.: Magnetism of finite graphene samples: Mean-field theory compared with exact diagonalization and quantum Monte Carlo simulations. {\it Phys. Rev. B} {\bf 2010}, 81, 115416.

\bibitem{18}
Feldner, H.; Meng, Z. Y.; Lang, T. C.; Assaad, F. F.; Wessel, S.; Honecker, A.: Dynamical Signatures of Edge-State Magnetism on Graphene Nanoribbons. {\it Phys. Rev. Lett.} {\bf 2011}, 106, 226401. 
	
\bibitem{19}
Mermin, N. D.; Wagner, H.: Absence of Ferromagnetism or Antiferromagnetism in One- or Two-Dimensional Isotropic Heisenberg Models. {\it Phys. Rev. Lett.} {\bf 1966}, 17, 1133−--1136.

\bibitem{20}
Yazyev, O. V.; Katsnelson, M. I.: Magnetic Correlations at Graphene Edges: Basis for Novel Spintronics Devices. {\it Phys. Rev. Lett.} {\bf 2008}, 100, 047209.

\bibitem{21}
Kane, C. L.; Mele, E. J.: Quantum Spin Hall Effect in Graphene. {\it Phys. Rev. Lett.} {\bf 2005}, 95, 226801.

\bibitem{22}
Moore, J. E.: The birth of topological insulators. {\it Nature} {\bf 2010}, 464, 194--−198.

\bibitem{23}
Hasan, M. Z.; Kane, C. L.: Colloquium: Topological insulators. {\it Rev. Mod. Phys.} {\bf 2010}, 82, 3045−--3067.

\bibitem{24}
Qi, X.-L.; Zhang, S.-C.: Topological insulators and superconductors. {\it Rev. Mod. Phys.} {\bf 2011}, 83, 1057−--1110.

\bibitem{25}
Gmitra, M.; Konschuh, S.; Ertler, C.; Ambrosch-Draxl, C.; Fabian, J.: Band-structure topologies of graphene: Spin-orbit coupling effects from first principles. {\it Phys. Rev. B} {\bf 2009}, 80, 235431.

\bibitem{26}
Weeks, C.; Hu, J.; Alicea, J.; Franz, M.; Wu, R.: Engineering a Robust Quantum Spin Hall State in Graphene via Adatom Deposition. {\it Phys. Rev. X} {\bf 2011}, 1, 021001.

\bibitem{27}
Yazyev, O. V.; Capaz, R. B.; Louie, S. G.: Theory of magnetic edge states in chiral graphene nanoribbons. {\it Phys. Rev. B} {\bf 2011}, 84, 115406.

\bibitem{28}
Akhmerov, A. R.; Beenakker, C. W. J.: Boundary conditions for Dirac fermions on a terminated honeycomb lattice. {\it Phys. Rev. B} {\bf 2008}, 77, 085423.

\bibitem{29}
Klein, D. J.; Bytautas, L.: Graphitic Edges and Unpaired $\pi$-Electron Spins. {\it J. Phys. Chem. A} {\bf 1999}, 103, 5196--−5210.

\bibitem{30}
Koskinen, P.; Malola, S.; H\"akkinen, H.: Self-Passivating Edge Reconstructions of Graphene. {\it Phys. Rev. Lett.} {\bf 2008}, 101, 115502.

\bibitem{31}
Girit, C. O.; Meyer, J. C.; Erni, R.; Rossell, M. D.; Kisielowski, C.; Yang, L.; Park, C.-H.; Crommie, M. F.; Cohen, M. L.; Louie, S. G.; Zettl, A.: Graphene at the Edge: Stability and Dynamics. {\it Science} {\bf 2009}, 323, 1705−--1708.

\bibitem{32}
Koskinen, P.; Malola, S.; H\"akkinen, H.: Evidence for graphene edges beyond zigzag and armchair. {\it Phys. Rev. B} {\bf 2009}, 80, 073401.

\bibitem{33}
Wassmann, T.; Seitsonen, A. P.; Saitta, A. M.; Lazzeri, M.; Mauri, F.: Structure, Stability, Edge States, and Aromaticity of Graphene Ribbons. {\it Phys. Rev. Lett.} {\bf 2008}, 101, 096402.

\bibitem{34}
Han, M. Y.; \"Ozyilmaz, B.; Zhang, Y.; Kim, P.: Energy Band-Gap Engineering of Graphene Nanoribbons. {\it Phys. Rev. Lett.} {\bf 2007}, 98, 206805.

\bibitem{35}
Bai, J.; Duan, X.; Huang, Y.: Rational Fabrication of Graphene Nanoribbons Using a Nanowire Etch Mask. {\it Nano Lett.} {\bf 2009}, 9, 2083−--2087.

\bibitem{36}
Ci, L.; Xu, Z.; Wang, L.; Gao, W.; Ding, F.; Kelly, K.; Yakobson, B.; Ajayan, P.: Controlled nanocutting of graphene. {\it Nano Res.} {\bf 2008}, 1, 116--−122.

\bibitem{37}
Campos, L. C.; Manfrinato, V. R.; Sanchez-Yamagishi, J. D.; Kong, J.; Jarillo-Herrero, P.: Anisotropic Etching and Nanoribbon Formation in Single-Layer Graphene. {\it Nano Lett.} {\bf 2009}, 9, 2600--−2604.

\bibitem{38}
Li, X.; Wang, X.; Zhang, L.; Lee, S.; Dai, H.: Chemically Derived, Ultrasmooth Graphene Nanoribbon Semiconductors. {\it Science} {\bf 2008}, 319, 1229--−1232.

\bibitem{39}
Wang, X.; Ouyang, Y.; Li, X.; Wang, H.; Guo, J.; Dai, H.: Room-Temperature All-Semiconducting Sub-10-nm Graphene Nanoribbon Field-Effect Transistors. {\it Phys. Rev. Lett.} {\bf 2008}, 100, 206803.

\bibitem{40}
Jiao, L.; Zhang, L.; Wang, X.; Diankov, G.; Dai, H.: Narrow graphene nanoribbons from carbon nanotubes. {\it Nature} 2009, 458, 877−--880.

\bibitem{41}
Kosynkin, D. V.; Higginbotham, A. L.; Sinitskii, A.; Lomeda, J. R.; Dimiev, A.; Price, B. K.; Tour, J. M.: Longitudinal unzipping of carbon nanotubes to form graphene nanoribbons. {\it Nature} {\bf 2009}, 458, 872−--876.

\bibitem{42}
Jiao, L.; Wang, X.; Diankov, G.; Wang, H.; Dai, H.: Facile synthesis of high-quality graphene nanoribbons. {\it Nature Nanotechnol.} {\bf 2010}, 5, 321--−325.

\bibitem{43}
Tao, C.; Jiao, L.; Yazyev, O. V.; Chen, Y.-C.; Feng, J.; Zhang, X.; Capaz, R. B.; Tour, J. M.; Zettl, A.; Louie, S. G.; Dai, H.; Crommie, M. F.: Spatially resolving edge states of chiral graphene nanoribbons. {\it Nature Phys.} {\bf 2011}, 7, 616−--620.

\bibitem{44}
Pan, M.; Gir\~ao, E. C.; Jia, X.; Bhaviripudi, S.; Li, Q.; Kong, J.; Meunier, V.; Dresselhaus, M. S.: Topographic and Spectroscopic Characterization of Electronic Edge States in CVD Grown Graphene Nanoribbons. {\it Nano Lett.} {\bf 2012}, 12, 1928−--1933.

\bibitem{45}
Zhang, X.; Yazyev, O. V.; Feng, J.; Xie, L.; Tao, C.; Chen, Y.-C.; Jiao, L.; Pedramrazi, Z.; Zettl, A.; Louie, S. G.; Dai, H.; Crommie, M. F.: Experimentally Controlling the Edge Termination of Graphene Nanoribbons. {\it ACS Nano} {\bf 2013}, 7, 198---202.

\bibitem{46}
Cai, J.; Ruffieux, P.; Jaafar, R.; Bieri, M.; Braun, T.; Blankenburg, S.; Muoth, M.; Seitsonen, A. P.; Saleh, M.; Feng, X.; M\"ullen, K.; Fasel, R.: Atomically precise bottom-up fabrication of graphene nanoribbons. {\it Nature} {\bf 2010}, 466, 470−--473.

\bibitem{47}
Blankenburg, S.; Cai, J.; Ruffieux, P.; Jaafar, R.; Passerone, D.; Feng, X.; M\"ullen, K.; Fasel, R.; Pignedoli, C. A.: Intraribbon Heterojunction Formation in Ultranarrow Graphene Nanoribbons. {\it ACS Nano} {\bf 2012}, 6, 2020−--2025.

\bibitem{48}
Ruffieux, P.; Cai, J.; Plumb, N. C.; Patthey, L.; Prezzi, D.; Ferretti, A.; Molinari, E.; Feng, X.; Müllen, K.; Pignedoli, C. A.; Fasel, R.: Electronic Structure of Atomically Precise Graphene Nanoribbons. {\it ACS Nano} {\bf 2012}, 6, 6930−--6935.

\bibitem{49}
Chuvilin, A.; Bichoutskaia, E.; Gimenez-Lopez, M. C.; Chamberlain, T. W.; Rance, G. A.; Kuganathan, N.; Biskupek, J.; Kaiser, U.; Khlobystov, A. N.: Self-assembly of a sulphur-terminated graphene nanoribbon within a single-walled carbon nanotube. {\it Nature Mater.} {\bf 2011}, 10, 687−--692.

\bibitem{50}
Chamberlain, T. W.; Biskupek, J.; Rance, G. A.; Chuvilin, A.; Alexander, T. J.; Bichoutskaia, E.; Kaiser, U.; Khlobystov, A. N.: Size, Structure, and Helical Twist of Graphene Nanoribbons Controlled by Confinement in Carbon Nanotubes. {\it ACS Nano} {\bf 2012}, 6, 3943−--3953.

\end{thebibliography}
\end{document}